# Bus Manufacturing Workshop Scheduling Method with Routing Buffer

Zhonghua Han, Jingyuan Zhang, Xiaoting Dong, Yuanwei Qi

*Abstract*—Aiming at solving the problem that the moving route is complicated and the scheduling is difficult in the routing buffer of the bus in the manufacturing workshop, a routing buffer mathematical programming model for bus manufacturing workshop is proposed. We design a moving approach for minimizing the total setup cost for moving in routing buffer. The framework and the solution ofthe optimization problem of such a bus manufacturing workshop scheduling with routing buffer are presented. The evaluation results show that, comparing with the irregularly guided moving method, the proposed method can better guide the bus movement in routing buffer by reducing the total setup time of all buses processed at the next stage, and obtaining a better scheduling optimization solution with minimize maximum total completion time.

## I. INTRODUCTION

With the features of multiple stages and parallel workstations, the painting production workshop of bus manufacturer isa typicalkind offlexible flowshop, whose scheduling problem has thetypical characteristics of flexible flowshop scheduling problem (FFSP) (Moslehiand Khorasanian, 2013), which is a typical NP-Hard problem. In recent years, there exist efficient heuristic methods for the classical FFSP (Han et al., 2012; Sun et al., 2017), FFSP with component altering times (Han et al., 2016), FFSP with multiple-objective optimization (Han et al., 2017), FFSP with Re-entrant problem (Han et al, 2018) and FFSP considering finite buffers (Han et al, 2018). In practice, the bus has a long production cycle, which requires a large number of buffer parking spaces to temporarily park the to-be-processedbuses. Because bus isa large-volume work-in-process, only a buffer with limited parking spaces can be installed on the productionline of bus. Therefore, the scheduling problem of bus manufacturing workshop belongs to the category of flexible flow shop limited buffer scheduling problem (Almederand Hartl, 2013). Different from the traditional flexible flow shop, there is a special limited buffer called routing buffer between two specific stages in the bus manufacturing workshop. Compared to the ordinary buffer and sequence buffer, the routing buffer not only has the parallel lanes for the to-be-processedbuses to move forward, but also provides electric flat carriages between parallel lanes.As a result, the bus can be moved forward in the lane,or moved in parallel between parallel lanes(Aref et al., 2012),which will lead to a more complicated movement for the bus in routing buffer. And due to the influence of routing buffer capacity, the bigger the capacity is, the greater the complexity of bus movement. It is necessary to consider the problem of buffer lane capacityand the optimization of bus movement in routing buffer, which further increases the difficulty of research problems.

The bus manufacturers adopt customized production method. There are a wide variety of models in production orders put into production every day. Customized production method brings differences in the processing time and process flow of the buses(Liu et al.,2008).At the two stages before and after routing buffer, processing time of diverse models is also different. Meanwhile, due to the limitation of workshop production resources, when a bus enters a workstation, it usually needs to implement setup operation whichis the preparationof the workstation for processing the next bus after it completes a bus processing, such as replacing the parts to be assembled, changing the color materials for spraying, exchanging the cutting and grinding tools, or cleaning and resetting the workstations during the processing gap. Therefore, the setup operation will result in a setup time in addition to the processing time of the workstation (Wang et al., 2010; Hakimzadeh and Zandieh, 2012). The setup time of the workstation will be affected by variousproperties such as the type, color, and shape of the processed object (semiconductor packaging line, automobile manufacturing and metallurgy industry) (Fu et al., 2012; Li et al., 2014),and also influenced by the material used in production and processing (semiconductor packaging line, assembly line of mechanical manufacturing, pharmaceutical industry, and tobacco industry)(Takano et al., 2017; Zhang et al., 2017). The replacement of cutting tools and molds for equipment will also add a setup time besides the standard processing time. Frequent setup operation will extend the bus production time, limiting company's existing production capacity(Thürer et al., 2013; Ding et al., 2016).

Reducing the impact of setup operation has a great effect on shortening the completion time of bus production. Solving the bushow to move more efficiently in routing buffer plays an important role in settling flexible flow shop scheduling problem and improving company's production benefit. As such, the research on scheduling problem of bus manufacturing workshop has important theoretical significance and application value. For the complex movementof to-be-processed bus in routing buffer and the setup operation of different buses at the next stage of routing buffer, this paper proposes the routing rules scheduling method with the minimum total setup cost(Fu et al.,2012; Davendra et al.,2013).By calculating the sumof setup cost of each moving path, this scheduling method establishes the routing rules to control the bus movement in routing buffer,so as to obtain the minimum total setup cost. Thus, it canguide the bus to move in routing buffer with the minimum setup cost,reduce the impact of setup operation, andsolve the scheduling problem that the bus movement in routing buffer of the manufacturing workshop is complex

## II. MATHEMATICAL MODEL OF ROUTING BUFFER PROBLEM

### A. Problem Description

The scheduling problem studied in this paper can be described as follows: As shown in Figure 1, a queue with $S$ busesis prepared to be processedat two stages before and after the routing buffer. At least one of these stagesincludes severalworkstations, and the busneeds to select a workstation for processing at each stage(El-Bouriand Nairy, 2011; Deng et al., 2016). There is a routing buffer with limited capacity between these two stages. After the to-be-processed bus enters the routing buffer after completing the previous stage, it will wait for entering the next stage. If the buffer is full, and there is no availableworkstation at the next stage, the to-be-processed buswill be stuck on the current stage workstation(Tasgetirenet al., 2015; Abdollahpourand Rezaeian, 2015). At this time, the workstation is blocked and it is impossible to process other buses.if the buffer has available space,the workstationwill be releasedand the bus will enter the buffer.The bus, passing through two stages before and after the routing buffer,adopts the first-in-first-out rule and continues to enter the follow-up stage for processing until the entire task is completed.The routing buffer consists of several parallel lanes with limited capacity. And the parking spaces in the parallel lanes can store the buses that are processed at previous stage and ready for the next stage. The electric flat carriages arranged between parallel lanescan transfer the to-be-processed bus from one buffer lane to anotheradjacent lane, so as to make the bus move between the lanes. In the buffer,the bus can choose to be moved forward in the lane, or moved in parallel between the lanes. It shows that the movingway has the characteristics of routing. The multiple lanes in the buffer and electric flat carriagesbetween the lanes constitute a routing network,where the bus movement is a process to select the path. When the bus is moved forward or movedin parallelin buffer lane, it will vacate its buffer parking space in the original buffer lane, so that other buses can enter this available space. Whenever a busis moved, the parking space of the bus in the buffer will be vacated,triggering the linkage of multiple buses.

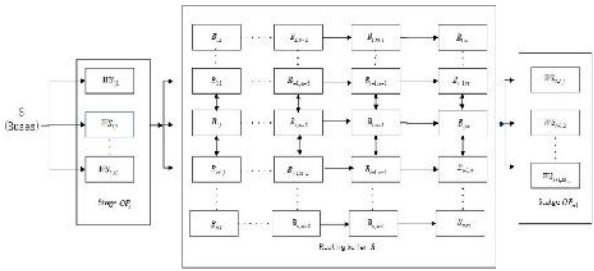

Figure .1.Movement rules of bus in routing buffer

### B. Model Parameters

The model parameters are sorted according to the first model element and then the evaluation indicator element.

$S$ indicates the total numberof buses to be processed.

$J_s$ means the to-be-processed bus, $s \in \{1,2,...,S\}$.

$B_{i,j}$ indicates the parking space in routing buffer, $i \in \{1,2,...,n\}$. $n$ represents the row number of parking space in routing buffer, $j \in \{1,2,...,m\}$. $m$ expresses the column number of routing buffer.

$O$ indicates the maximum amount of setup factor.

$CK_x$ means varioussetup factors, such as exchanging color materials, cleaning tools, replacing parts, etc. $x \in \{1,2,...,O\}$.

$J\_CK_{s,x}$ represents the setup factor $CK_x$ of bus $J_s$.

$Y_x$ indicates the maximum amount of bus setup factor property.

$T\_JTyco_{x,y,y'}$ means the setup time of bus with different setup factor's different property, the setup time $y, y' \in \{1,2,...,Y_x\}$.

$Z$ represents the maximum amount of bus model.

$Type_y$ means model property of the bus. $z \in \{1,2,...,Z\}$.

$Color_y$ means color property of the bus.

$P$ indicates the maximum amount of bus type. A type of bus with a specific model and color is defined as the bus type.

$Typei$ represents the type of to-be-processed bus, $Typei \in \{1,2,...,P\}$. (A type of bus that includes properties such as a certain type of vehicle and a certain color.)

$J\_C_{s,Typei}$ represents thetype of bus $J_s$.

$T\_JC_{Typei,Typei'}$ expresses the setup cost of different types of bus successively-processed at the same workstation, setup cost $Typei, Typei' \in \{1,2,...,p\}$.

$J\_CD_{s,s',x}$ indicates whether the setup factors x of two different types of buses successively-processed at the same workstation are the same.

$T\_CD_{s,s'}$ denotes the total setup time for two different types of buses successively-processed at the same workstation, $s, s' \in \{1,2,...,S\}$.

$T\_BJ_{(i,j),(i',j')}$ represents the setup cost between buffer parking space $B_{i,j}$ and its adjacent parking space, $|i-i'| \leq 1$, $|j-j'| \leq 1$, and $|i-i'|+|j-j'|=1$.

$J\_BQ_{i,j}$ indicates whether there is a to-be-processed bus on the parking space $B_{i,j}$ in routing buffer.

$TSum\_BJ$ denotes the total setup cost after a linkage transfer of the buffer due to the bus online processing in the buffer.

$Q$ represents the maximum amount of processing stage in the bus manufacturing workshop.

$OP_l$ means the processing stagein the bus manufacturing workshop, $l \in \{1,2,...,Q\}$.

$M_l$ indicates the maximum amount of workstations of stage $OP_l$.

$WS_{l,t}$ indicates the $t$ th workstation of stage $OP_l$, $t \in \{1, 2, ..., M_l\}$.

$A_{l,t}$ indicates the setup frequency of the workstation $WS_{l,t}$ at stage $OP_l$ after completing all buses in this workstation, $A_{l,t} \in \{1, 2, 3, ..., S-1\}$.

$J\_WS_{l,t,s,s'}$ denotes whether the buses $J_s$ and $J_{s'}$ have setup operation at workstation $WS_{l,t}$.

$T\_Sum_{l,t}$ represents the total setup time of workstation $WS_{l,t}$ at stage $OP_l$ after completing all buses at this workstation, $T\_Sum_{l,t} = \sum_{s=1}^{S-1} \sum_{s'=s+1}^{S} T\_CD_{s,s'} \cdot J\_WS_{l,t,s,s'}$

$St_{s,l,t}$ means the start processing time of to-be-processed bus $J_s$ at workstation $WS_{l,t}$ at stage $OP_l$, $St_{s,l,t} \geq 0$.

$En_{s,l,t}$ means the processing completion time of to-be-processed bus $J_s$ at workstation $WS_{l,t}$ at stage $OP_l$.

$Tw_{s,l,t}$ means the processing time of to-be-processed bus $J_s$ at workstation $WS_{l,t}$ at stage $OP_l$.

$WSend_{BuWP}$ indicates the time when all buses are finished the processing at the next stage $OP_{BuWP}$ of routing buffer.

$TWSfinal$ indicates the completion time of all buses processed at the last stage.

*C. 2.3 Constraints*

(1) Uninterruptible constraint: Each bus cannot be interrupted once it starts processing on the equipment until the bus is completed.
(2) Equipment uniqueness constraint: Each parallel workstation can only process one bus at the same time
(3) Equipment availability constraint: Allow waiting between stages, and permit workstation to be idle when the to-be-processed bus is not arriving.
(4) Buffer constraint: Not consider time consumption of the bus moving in the buffer.

$$J\_BQ_{i,j} = \begin{cases} 1 & \text{There is to-be-processed bus at packing space } B_{i,j} \\ 0 & \text{There isn't to-be-processed bus at packing space } B_{i,j} \end{cases}$$

$$J\_CD_{s,s',x} = \begin{cases} 1 & \text{Buses } J_s \text{ and } J_{s'} \text{ with the same setup factor } CK_x, \text{ occurring setup operation} \\ 0 & \text{Buses } J_s \text{ and } J_{s'} \text{ with different setup factor } CK_x, \text{ not occurring setup operation} \end{cases}$$

$$J\_WS_{l,t,s,s'} = \begin{cases} 1 & \text{Buses } J_s \text{ and } J_{s'} \text{ occur setup operation at workstation } WS_{l,t} \\ 0 & \text{Buses } J_s \text{ and } J_{s'} \text{ not occur setup operation at workstation } WS_{l,t} \end{cases}$$

$$En_{s,l,t} = St_{s,l,t} + Tw_{s,l,t}, s \in \{1, 2, ..., S\}, l \in \{1, 2, ..., Q\}, \quad (1)$$
$$t \in \{1, 2, ..., M_l\}$$

Equation (1) indicates that the processing completion time of the to-be-processed bus $J_s$ at workstation $WS_{l,t}$ of stage $OP_l$ is equal to the sum of the start processing time and processing time at this workstation.

$$En_{s,l,t} > St_{s,l,t}, s \in \{1, 2, ..., S\}, l \in \{1, 2, ..., Q\}, \quad (2)$$
$$t \in \{1, 2, ..., M_l\}$$

Equation (2) indicates that the processing completion time of the to-be-processed bus $J_s$ at workstation $WS_{l,t}$ of stage $OP_l$ is greater than the start processing time at this workstation.

$$St_{s,l+1,t} \geq En_{s,l,t}, s \in \{1, 2, ..., S\}, l \in \{1, 2, ..., Q\}, \quad (3)$$
$$t \in \{1, 2, ..., M_l\}$$

Equation (3) indicates that the start processing time of the to-be-processed bus $J_s$ at workstation $WS_{l+1,t}$ of stage $OP_{l+1}$ is greater than or equal to the processing completion time at workstation $WS_{l,t}$ of stage $OP_l$.

$$TWSfinal = \max(En_{s,Q,t}) \quad (4)$$

Equation (4) indicates that the completion time of the processing task is equal to the processing completion time of the last bus at the last stage.

## III. ALGORITHM PROCESS

The setup cost refers to the cost relationship between the parking space and its adjacent parking space in routing buffer, which means that if two buses on two adjacent parking spaces are processed at the same workstation, their setup time will be the same. It shows the degree of similarity in processing requirements between two buses, the more similar the processing requirements, the less setup time for them to be processed at the same workstation, and the smaller the setup cost between the parking spaces. The setup cost is:

$$T\_JC_{Typei,Typei'} = T\_CD_{s,s'} = \sum_{x=1}^{O} T\_JTyco_{x,y,y'} \cdot J\_CD_{s,s',x} \quad (5)$$

The setup cost between two adjacent parking spaces is dynamically changing, with the changes in the processing requirements of buses stored on it. The total setup cost refers to the sum of setup costs of all to-be-processed buses and their adjacent to-be-processed buses in the buffer after the completion of a bus linkage process:

$$TSum\_BJ = \sum_{i=1}^{n}\sum_{j=1}^{m} T\_BJ_{(i,j),(i',j')} \cdot J\_BQ_{i,j} \cdot J\_BQ_{i',j'}, \quad (6)$$
$$|i-i'| \leq 1, |j-j'| \leq 1, and\ |i-i'|+|j-j'|=1$$

The moving way of the bus in routing buffer is proposed in accordance with the complex moving path of the bus in routing buffer and two stages before and after routing buffer with setup operation. That is, when the bus in the buffer leaves the current parking space to be processed, it will trigger the linkage of the bus in the buffer. By calculating the total setup cost of all possible linkage processes, we find and choose the moving path with the minimum total setup cost, so as to construct the moving way of the bus in routing buffer with the minimum total setup cost, which can reduce the setup time and frequency of the bus at the next stage. This moving way of the bus in routing buffer can guide the bus in routing buffer to move with the minimum setup cost during the entire process. As such, the scheduling scheme for bus manufacturing workshop with routing buffer is given, to make the bus with similar processing requirements processed at the same workstation, improve processing efficiency, and develop company's existing production capacity and its benefit. Specific steps are as follows:

Step 1: Establish the initial population, generating an online individual $p_{rand(1,S)}$ randomly. The to-be-processed bus is processed in accordance with the online sequence.

Step2: Determine whether a bus $J_s$ in buffer parking space $B_{i,j}$ is processed at buffer's next stage $OP_{l+1}$. If not, continue to wait to-be-processed buses to enter the buffer. If so, execute Step 3.

Step3: Determine whether there is a to-be-processed bus on adjacent parking space of space $B_{i,j}$ storing bus $J_s$. If not, the linkage process is completed. If so, repeat Step 3.

Step4: Calculate the total setup cost $TSum\_BJ$ of the to-be-processed bus in the buffer after a move.

Step5: Determine whether all linkage situations are completed. If not, repeat Step3. If so, compare the total setup costs of all linkage situations to obtain the minimum value.

Step6: Transfer the to-be-processed bus in routing buffer with the minimum total setup cost to complete the movement process.

Step7: Determine whether the processing task is completed. If so, it will end. If not, repeat Step2.

The flow chart is as follows:

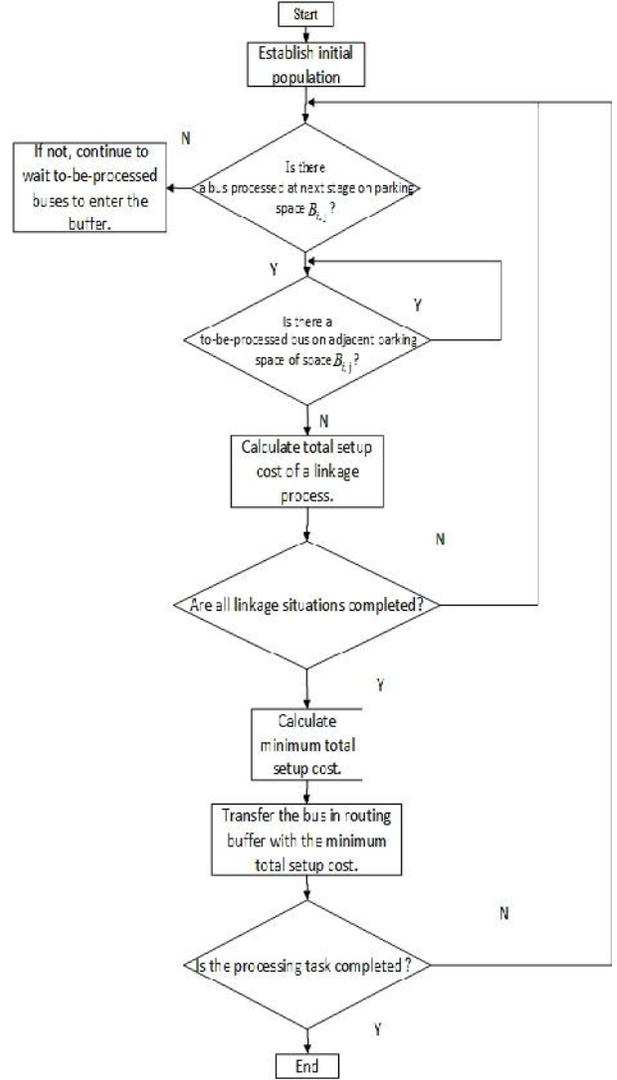

Figure .2. Flow chart of setup cost rules

## IV. EXPERIMENTAL AND SIMULATION

### A. Establishing Simulation Data

(1) Workshop model parameters

Scheduling simulation data includes four stages, namely, $\{OP_1, OP_2, OP_3, OP_4\}$. The number of parallel workstations $\{M_l\}$ for the four workstations is $\{3,2,3,3\}$. The buffer between the welding workshop and painting workshop is routing buffer, so the buffer between stages $OP_1$ and $OP_2$ in scheduling data is set to the routing buffer. In routing buffer, the row number is $n=4$, while the column number is $m=3$. During the production process in painting workshop, if the model and color properties of two successively-processed buses take changes, it needs to clean the workstation and adjust its production equipment. As a result, the simulation process employs the changes in bus model and color properties as the basis for calculating the preparation time. From Table 1, it shows the preparation time parameters when the model and color of the buses processed successively on workstation take changes.

TABLE I MODEL PARAMETERS

| Model parameters | | Parameter description | Parameter value (unit: min) |
|---|---|---|---|
| Related parameters of limited buffer | $n$ | Row number of routing buffer | 4 |
| | $m$ | Column number of routing buffer | 3 |
| Related parameters of preparation time | $CK_1$ | On the parallel workstation of stage $OP_2$, the preparation time parameters of successively-processed bus models changing from $Type_1$ to $Type_2$. | 10 |
| | | On the parallel workstation of stage $OP_2$, the preparation time parameters of successively-processed bus models changing from $Type_1$ to $Type_3$. | 4 |
| | | On the parallel workstation of stage $OP_2$, the preparation time parameters of successively-processed bus models changing from $Type_2$ to $Type_3$. | 14 |
| | $CK_2$ | On the parallel workstation of stage $OP_2$, the preparation time parameters of successively-processed bus color changing from $Color_1$ to $Color_2$. | 13 |
| | | On the parallel workstation of stage $OP_2$, the preparation time parameters of successively-processed bus color changing from $Color_1$ to $Color_3$. | 17 |
| | | On the parallel workstation of stage $OP_2$, the preparation time parameters of successively-processed bus color changing from $Color_2$ to $Color_3$. | 12 |

(2) Parameters of Processed Object

The simulation data comes from the actual production data of a bus company.

The sum of bus properties is $O = 2$. $CK_1$ represents the model property of the bus, while $CK_2$ means the color property of the bus. The value of model property is $CK_1 = \{Type_1, Type_2, Type_3\}$, while the value of color property is $CK_2 = \{Color_1, Color_2, Color_3\}$. Suppose that two successively-processed buses on workstation $WS_{2,2}$ of stage $OP_2$ are bus $J_3$ and $J_5$, if properties of bus $J_3$ are $CK_1 = Type_3$, $CK_2 = Color_2$, and properties of bus $J_5$ are $CK_1 = Type_1$, $CK_2 = Color_3$, then $J\_CK_{3,1} \neq J\_CK_{5,1}$, $J\_CK_{3,2} \neq J\_CK_{5,2}$. From Table 3, we can conclude $T\_CD_{3,5} = 16$.

TABLE II INFORMATION OF BUS MODEL AND COLOR PROPERTIES

| Bus properties | Model | Color |
|---|---|---|
| Bus $J_1$ | $Type_1$ | $Color_1$ |
| Bus $J_2$ | $Type_3$ | $Color_2$ |
| Bus $J_3$ | $Type_2$ | $Color_1$ |
| Bus $J_4$ | $Type_2$ | $Color_1$ |
| Bus $J_5$ | $Type_1$ | $Color_1$ |
| Bus $J_6$ | $Type_2$ | $Color_2$ |
| Bus $J_7$ | $Type_1$ | $Color_3$ |
| Bus $J_8$ | $Type_1$ | $Color_1$ |
| Bus $J_9$ | $Type_2$ | $Color_2$ |
| Bus $J_{10}$ | $Type_3$ | $Color_2$ |
| Bus $J_{11}$ | $Type_1$ | $Color_3$ |
| Bus $J_{12}$ | $Type_1$ | $Color_3$ |
| Bus $J_{13}$ | $Type_3$ | $Color_2$ |
| Bus $J_{14}$ | $Type_2$ | $Color_2$ |
| Bus $J_{15}$ | $Type_2$ | $Color_1$ |
| Bus $J_{16}$ | $Type_1$ | $Color_1$ |
| Bus $J_{17}$ | $Type_1$ | $Color_3$ |
| Bus $J_{18}$ | $Type_3$ | $Color_2$ |
| Bus $J_{19}$ | $Type_2$ | $Color_1$ |
| Bus $J_{20}$ | $Type_2$ | $Color_2$ |
| Bus $J_{21}$ | $Type_3$ | $Color_2$ |
| Bus $J_{22}$ | $Type_2$ | $Color_1$ |

TABLE III SCHEDULE OF DIVERSE SETUP

| | $Type_1$ $Color_1$ | $Type_2$ $Color_1$ | $Type_3$ $Color_2$ | $Type_2$ $Color_2$ | $Type_1$ $Color_3$ |
|---|---|---|---|---|---|
| $Type_1$ $Color_1$ | 0 | 6 | 16 | 14 | 9 |
| $Type_2$ $Color_1$ | 6 | 0 | 16 | 8 | 15 |
| $Type_3$ $Color_2$ | 16 | 16 | 0 | 8 | 16 |
| $Type_2$ $Color_2$ | 14 | 8 | 8 | 0 | 14 |
| $Type_1$ $Color_3$ | 9 | 15 | 16 | 14 | 0 |

TABLE IV STANDARD PROCESSING HOURS FOR BUS PRODUCTION

| | $OP_1$ | $OP_2$ | $OP_3$ | $OP_4$ |
|---|---|---|---|---|
| $J_1$ | 8 | 30 | 34 | 42 |
| $J_2$ | 11 | 38 | 38 | 36 |
| $J_3$ | 15 | 28 | 44 | 26 |
| $J_4$ | 19 | 25 | 42 | 24 |
| $J_5$ | 10 | 26 | 52 | 34 |
| $J_6$ | 16 | 36 | 40 | 30 |
| $J_7$ | 12 | 20 | 46 | 28 |
| $J_8$ | 21 | 24 | 48 | 32 |
| $J_9$ | 22 | 22 | 35 | 38 |
| $J_{10}$ | 13 | 32 | 36 | 40 |
| $J_{11}$ | 20 | 35 | 45 | 44 |
| $J_{12}$ | 14 | 34 | 50 | 22 |
| $J_{13}$ | 8 | 30 | 34 | 42 |
| $J_{14}$ | 11 | 38 | 38 | 36 |
| $J_{15}$ | 15 | 28 | 44 | 26 |
| $J_{16}$ | 19 | 25 | 42 | 24 |
| $J_{17}$ | 10 | 26 | 52 | 34 |
| $J_{18}$ | 16 | 36 | 40 | 30 |
| $J_{19}$ | 12 | 20 | 46 | 28 |
| $J_{20}$ | 21 | 24 | 48 | 32 |
| $J_{21}$ | 22 | 22 | 35 | 38 |
| $J_{22}$ | 13 | 32 | 36 | 40 |

*B. Simulation Scheme*

The moving way of the bus in routing buffer based on total setup cost as evaluation index is compared with buffer movement scheme without optimization, and the total setup time $T\_Sum_{l,t}$ processed on each workstation at buffer's next stage is analyzed. Comparing two completion times of these two schemes after completing the processing task, namely, the completion time $WSend_{BuWP}$ of routing buffer's next stage and the completion time $TWSfinal$ of the processing task. Analyzing the optimization effect on moving way of the bus in routing buffer based on total setup cost as evaluation index. In addition, each program implements 20 times independently.

TABLE V SIMULATION INFORMATION OF TWO SCHEMES

| Simulation schemes | Algorithm parameters |
|---|---|
| Scheme 1 | The moving way of the bus in routing buffer based on total setup cost as evaluation index |
| Scheme 2 | Buffer movement scheme with randomly moving |

*C. Simulation Results and Analysis*

The moving way of the bus in routing buffer based on total setup cost as evaluation index is implemented by the MATLAB2012b simulation software, which is running on the Windows 10 operating system, Core i7 processor, CPU2.20GHz, and the PC with 8GB memory.

(1) Evaluation index of scheduling results

TABLE VI RESULTS OF TWO SIMULATION SCHEMES

| Evaluation index | | Scheme 1 | Scheme 2 |
|---|---|---|---|
| $T\_Sum_{BuWP,1}$ | Optimal value | 13.0 | 26.0 |
| | Worst value | 98.0 | 88.0 |
| | Average value | 49.8 | 58.0 |
| $T\_Sum_{BuWP,2}$ | Optimal value | 23.0 | 37.0 |
| | Worst value | 92.0 | 86.0 |
| | Average value | 33.9 | 59.5 |
| $WSend_{BuWP}$ | Optimal value | 350.0 | 355.0 |
| | Worst value | 387.0 | 413.0 |
| | Average value | 369.6 | 385.0 |
| $TWSfinal$ | Optimal value | 401.0 | 407.0 |
| | Worst value | 451.0 | 386.0 |
| | Average value | 423.0 | 431.7 |

TABLE VII SUMMARY OF OPTIMIZATION RESULTS

| | Scheme 1 (Average value) | Scheme 2 (Average value) | Optimization range | Optimization ratio |
|---|---|---|---|---|
| $T\_Sum_{BuWP,l}$ | 41.8 | 58.7 | 16.9 | 28.8% |
| $WSend_{BuWP}$ | 369.6 | 385.0 | 15.4 | 4% |
| $TWSfinal$ | 423.0 | 431.7 | 8.7 | 2% |

From Table 6 and Table 7, it can be concluded that the average total setup time for routing buffer's next stages

$WS_{BuWP,1}$ and $WS_{BuWP,2}$ are reduced by 8.2 and 25.6 respectively. The average total setup time at this stage is reduced by 16.9, with 28.8% optimization ratio, while the completion time at this stage is reduced by 15.4, with 4.0% optimization ratio. The completion time for processing the task is reduced by 8.7, with 2.0% optimization ratio. It can be seen that the moving way of the bus in routing buffer based on total setup cost as evaluation index can better guide the bus to move in routing buffer, which can reduce the total setup time of all buses processed at the next stage, and also decrease the completion time of to-be-processed buses at routing buffer's next stage and the minimum and maximum total completion time in the global scheduling process.

(2) Gantt graph analysis of movement process and scheduling results in routing buffer

Figure 3 (a) shows the distribution of to-be-processed buses in routing buffer at a moment. Figure 3 (b) shows the result of bus movement guided by the moving way of the bus in routing buffer based on total setup cost as evaluation index. Figure 3 (c) shows the result of the movement without using any method. It can be seen from the figure that at this moment, the buses in buffer trigger the linkage because of the on-line processing of the bus on parking space $B_{1,3}$. The linkage operation of Figure 3 (b) is: $J_3 \to B_{1,3}$, $J_{15} \to B_{2,3}$, $J_{20} \to B_{2,2}$. The linkage process of Figure 3 (c) is: $J_{11} \to B_{1,3}$, $J_9 \to B_{1,2}$, $J_{20} \to B_{1,1}$, $J_{18} \to B_{2,1}$, $J_{10} \to B_{3,1}$. The total setup cost of Figure 3 (b) is $TSum\_BJ$ =134, and that of Figure 3(c) is $TSum\_BJ'$ =140.

By adopting Scheme 1 and Scheme 2, the scheduling results of Gantt graphs are shown in Figure 4 (a) and Figure 4 (b) respectively. The abscissa is the time axis, while the ordinate indicates the workstation for each stage. The black part indicates the situation of the bus processed at the workstation without setup operation, while the red part denotes the situation of the bus processed at the workstation with setup operation; the green part represents the movement of the to-be-processed bus in routing buffer. It can be seen from the figure that the stage with setup rules in this simulation is the second stage. After the processing at the first stage, the to-be-processed bus enters the buffer, taking a movement according to two schemes' rules and then entering the next stage for on-line processing. The green part in the Gantt graph is the movement of the bus in the buffer with time changes. In the Gantt, the gap in processing time represents the setup time between two buses successively processed at this workstation. It can be seen from Figure 4 (a) that the setup frequency at routing buffer's next stage using scheme 1 is 6 times, the average value of setup time is 34, the completion time at this stage is 354, and the completion time for this task is 400. From Figure 4 (b), it can be concluded that the setup frequency at routing buffer's next stage using scheme2 is 8 times, the average value of setup time is 78, the completion time at this stage is 397, and the completion time for this processing task is 440. Therefore, the moving way of the bus in routing buffer based on total setup cost as evaluation index can better guide the bus to move in routing buffer, which can reduce the total setup time of all buses processed at the next stage, and also decrease the minimum and maximum total completion time in the global scheduling process.

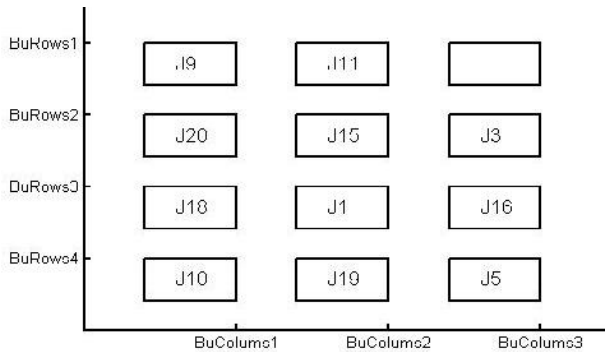

Figure 3 (a)

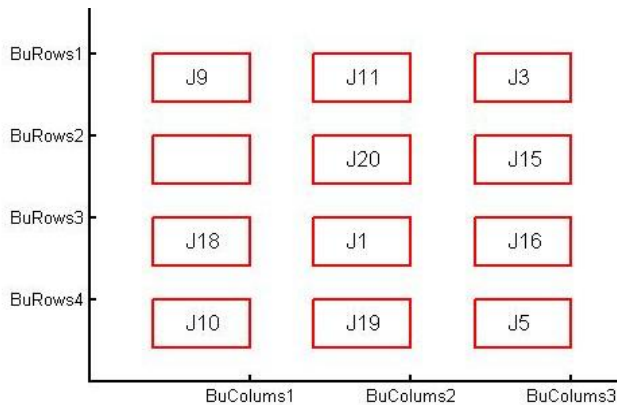

Figure 3 (b)

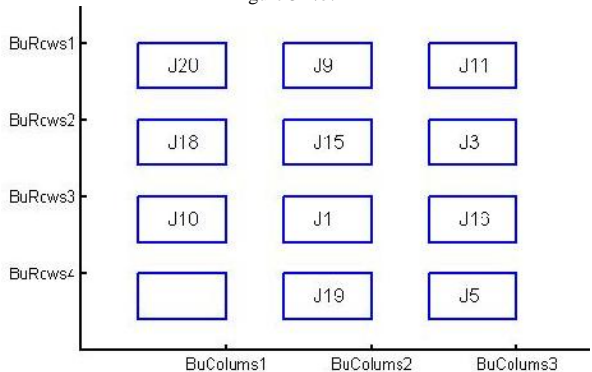

Figure 3 (c)

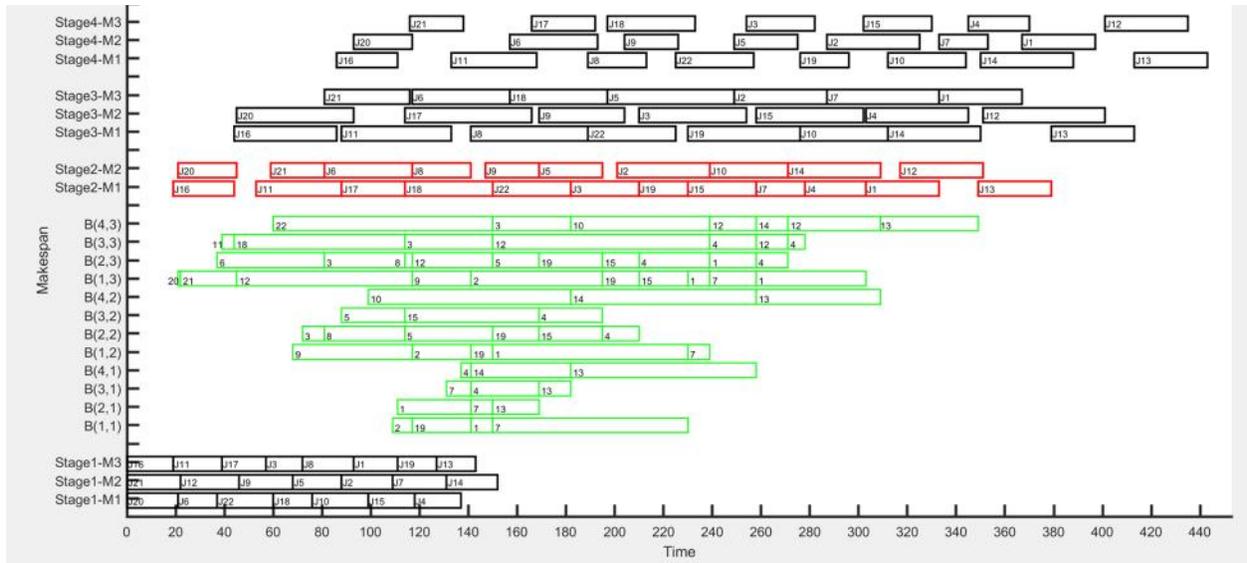

Figure 4(a) The simulation Gantt graph of scheme 1

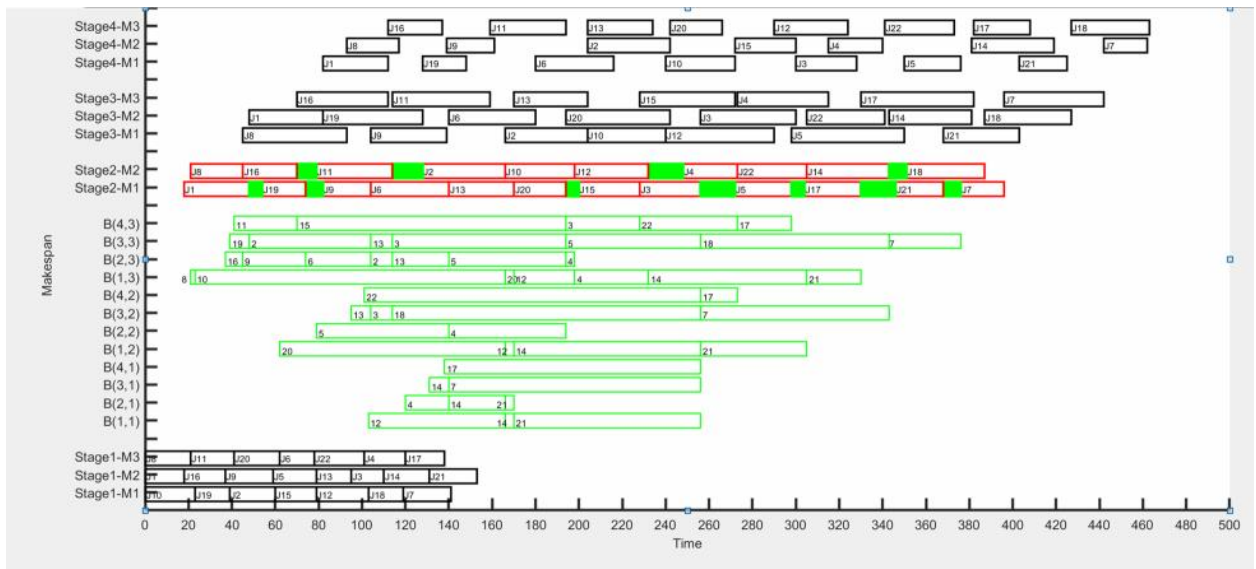

Figure 4(b) The simulation Gantt graph of scheme 2

## V. CONCLUSION

In this paper, we study a complexbus manufacturing workshop scheduling problem with routing buffer and setup operation. Because the routing buffer has multiple parallel lanes and there are electric flat carriages between the lanes, both of them make the bus movement in the buffer become very complicated, which has the characteristics of routing. According to the complex movement of to-be-processed bus in routing buffer and the setup operation at the next stage, this article proposes a moving way of the bus in routing buffer based on total setup cost as evaluation index,to guide the bus to move in routing buffer with the minimum total setup cost.As such, a scheduling scheme for bus manufacturing
workshop with routing buffer is given, so as to solve the scheduling problem that the bus movement in routing buffer of the manufacturing workshop is complex and difficult to find the optimal way.Through comparing the moving way of the bus in routing buffer based on total setup cost as evaluation index with buffer movement scheme without optimization, it can be concluded that the moving way of the bus in routing buffer based on total setup cost as evaluation index can better guide the bus to move in routing buffer, which can reduce the total setup time of all buses processed at the next stage, and also decrease the minimum and maximum total completion time in the global scheduling process.

**Acknowledgement:** This work was supported by Liaoning Provincial Science Foundation, China(No. 2018106008)，Project of Liaoning ProvinceEducation Department, China (No. LJZ2017015) and Shenyang Municipal Science and Technology Project, China (No. Z18-5-015).

REFERENCES

Ghasem Moslehi and Danial Khorasanian (2013) 'A hybrid variable neighborhood search algorithm for solving the limited-buffer permutation flow shop scheduling problem with the makespan criterion', *Computers and Operations Research,* pp.2-7.

Zhonghua Han, Xiaofu Ma, Lili Yao and Haibo Shi. (2012) 'Cost Optimization Problem of Hybrid Flow-Shop Based on PSO Algorithm', *Advanced Materials Research*, Vol. 532, pp. 1616-1620.

Yue Sun, Shuo Lin, Tan Li and Xiaofu Ma. (2017) 'Improved imperialist competitive algorithm for flexible flow shop scheduling', 2017 9th International Conference on Modelling, Identification and Control (ICMIC), Kunming, pp. 169-174.

Zhonghua Han, Yihang Zhu, Xiaofu Ma and Zhili Chen. (2016) 'Multiple rules with game theoretic analysis for flexible flow shop scheduling problem with component altering times', International Journal of Modelling, Identification and Control, Vol.26, No.1, pp.1-18.

Zhonghua Han, Shiyao Wang, Xiaoting Dong and Xiaofu Ma. (2017) 'Improved NSGA-II algorithm for multi-objective scheduling problem in hybrid flow shop', 2017 9th International Conference on Modelling, Identification and Control (ICMIC), Kunming, pp. 740-745.

Zhonghua Han, Xutian Tian, Xiaofu Ma and Shuo Lin. (2018) 'Scheduling for Re-entrant Hybrid Flowshop Based on Wolf Pack Algorithm', IOP Conference Series: Materials Science and Engineering, Vol. 382, No. 3.

Zhonghua Han, Yue Sun, Xiaofu Ma and Zhe Lv. (2018) 'Hybrid flow shop scheduling with finite buffers', International Journal of Simulation and Process Modelling, Vol.13, No.2, pp. 156-166.

C. Almeder and R. F. Hartl. (2013) 'A metaheuristic optimization approach for a real-world stochastic flexible flow shop problem with limited buffer', International Journal of Production Economics, 145(1). pp.88-95.

Aref Maleki-Darounkolaei, Mahmoud Modiri, Reza Tavakkoli-Moghaddam and Iman Seyyedi. (2012) 'A three-stage assembly flow shop scheduling problem with blocking and sequence-dependent set up times', *Journal of Industrial Engineering International*, 8(1). pp.26.

Bo Liu, Ling Wang and Yi-Hui Jin. (2008) 'An effective hybrid PSO-based algorithm for flow shop scheduling with limited buffers', *Computers & Operations Research*, 37(1). pp. 18-27.

Ling Wang, Quan-Ke Pan, P. N. Suganthan, Wen-Hong Wang and Ya-Min Wang. (2010) 'A novel hybrid discrete differential evolutional algorithm for blocking flow shop scheduling problems',*Computers & Operations Research*, 37(3). pp. 509-520.

Sina Hakimzadeh and Abyaneh M. Zandieh. (2012) 'Bi-objective hybrid flow shop scheduling with sequence-dependent setup times and limited buffers', *Int J Adv Manuf Technol*, 58(1-4). pp. 309–325.

Qing Fu, AppaIyer Sivakumar and KunpengLi. (2012) 'Optimisation of flow-shop scheduling with batch processor and limited buffer', *International Journal of Production Research*, pp. 2267–2285.

Jun-qing Li and Quan-ke Pan. (2014) 'Solving the large-scale hybrid flow shop scheduling problem with limited buffers by a hybrid artificial bee colony algorithm', *Information Sciences*, 316(c). pp. 487-502.

Cheng Zhang, Zhongshun Shi, Zewen Huang, Yifan Wu and Leyuan Shi. (2017) 'Flow shop scheduling with a batch processor and limited buffer', International Journal of Production Research, 55(11). pp. 3217-3233.

Mauricio Iwama Takano, Marcelo Seido Nagano and Wenjun Xu. (2017) 'A branch-and-bound method to minimize the makespan in a permutation flow shop with blocking and setup times', *Cogent Engineering*. 4(1).

Matthias Thürer, MoacirGodinho Filho and Mark Stevenson. (2013) 'Coping with finite storage space in job shops through order release control: an assessment by simulation', *International Journal of Computer Integrated Manufacturing*, 26(9). pp. 830-838.

Jian-Ya Ding, Shiji Song, Jatinder N.D. Gupta, Cheng Wang, Rui Zhang and Cheng Wu. (2016) 'New block properties for flowshop scheduling with blocking and their application in an iterated greedy algorithm', *International Journal of Production Research*, 54(16). pp. 1-14.

Qing Fu, AppaIyer Sivakumar and Kunpeng Li. (2012) 'Optimisation of flow-shop scheduling with batch processor and limited buffer', *International Journal of Production Research*, pp. 2267-2285.

Donald Davendra and Magdalena Bialic-Davendra. (2013) 'Scheduling flow shops with blocking using a discrete self-organising migrating algorithm',*International Journal of Production Research*, 51(8). pp. 2200-2218.

Ahmed El-Bouri and Subrahmanya Nairy. (2011) 'An investigation of cooperative dispatching for minimising mean flowtime in a finite-buffer-capacity dynamic flowshop', *International Journal of Production Research*, 49(6). pp. 1785-1800.

Guanlong Deng, Hongyong Yang and Shuning Zhang. (2016) 'An Enhanced Discrete Artificial Bee Colony Algorithm to Minimize the Total Flow Time in Permutation Flow Shop Scheduling with Limited Buffers', *Mathematical Problems in Engineering*, pp. 1-11.

M. Fatih Tasgetiren, Quan-Ke Pan, Damla Kizilay and Gursel Suer. (2015) 'A populated local search with differential evolution for blocking flowshop scheduling problem', *IEEE Congress on Evolutionary Computation (CEC)*, pp. 2789-2796.

Sana Abdollahpour and Javad Rezaeian. (2015) 'Minimizing makespan for flow shop scheduling problem with intermediate buffers by using hybrid approach of artificial immune system', *Applied Soft Computing*, pp. 44-56.